\documentclass[prl,twocolumn,showpacs]{revtex4}

\usepackage{amsmath}
\usepackage{amssymb}
\usepackage{graphicx}

\newcommand{\uinvnorm}{|\kern-2pt|\kern-2pt|}

\begin{document}
\bibliographystyle{apsrev}

\title{The Propagation of Quantum Information Through a Spin System}

\author{Tobias J.\ Osborne}
\email[]{T.J.Osborne@bristol.ac.uk}
\author{Noah Linden}
\email[]{N.Linden@bristol.ac.uk} \affiliation{Department of
Mathematics, University of Bristol, University Walk, Bristol BS8
1TW, United Kingdom}
\date{19th November 2003}

\begin{abstract}
It has been recently suggested that the dynamics of a quantum spin
system may provide a natural mechanism for transporting quantum
information.  We show that one dimensional rings of qubits with
fixed (time-independent) interactions, constant around the ring,
allow high fidelity communication of quantum states.  We show that
the problem of maximising the fidelity of the quantum
communication is related to a classical problem in fourier wave
analysis. By making use of this observation we find that if both
communicating parties have access to limited numbers of qubits in
the ring (a fraction that vanishes in the limit of large rings) it
is possible to make the communication arbitrarily good.
\end{abstract}


\maketitle

In quantum information science it is a crucial problem to develop
techniques for communicating qubits. A key task is to develop
protocols for taking a qubit's state at one location to another
while minimising the degradation of the quantum coherence.

In this paper we consider whether it is possible to use a one
dimensional  arrangement of qubits -- a ring -- coupled by
nearest-neighbour interactions, to communicate a qubit from one
part of the ring to another with high-fidelity.

It is clear that this communication can be done, for example, if
one can perform fast local unitary operations on the individual
qubits (by fast we mean that the local unitary operations are
effectively instantaneous compared with the time it takes the
interaction between qubits to change the state appreciably).  This
is because any nearest-neighbour interaction can be used to
perform a swap operation on pairs of qubits \cite{duer:2001a,
bennett:2002a, dodd:2002a, khaneja:2002a}. Thus the communication
can be performed by putting a given qubit into the state to be
communicated and then moving it along the chain of qubits by
swapping the state along the line to the required position.

In this paper, however, we have in mind the situation that we do
not have access to the whole set of spins during the
communication; rather the set of spins behaves more like a fibre
into which one can input a qubit state, and hope that the state
appears at the other end of the fibre.  The question is whether it
is possible for a set of spins with only its {\em passive}
interaction to act as an effective communication channel.

We would also like the qubit fibre to have the property that we
could increase the length of the fibre without needing to
completely re-fabricate it; thus if we wanted to communicate twice
as far we would like to be able to simply add two identical fibres
together and find that the fidelity and time of transmission
depended in a simple way on the fidelity and time of the
individual fibre.

Some interesting work has already been done in this area
concerning communication of qubits along a 1D Heisenberg spin
chain \cite{bose:2002a} (and, more recently, in the $1$D $XY$
model, \cite{subrahmanyam:2003a}). In \cite{bose:2002a} the aim
was to put one qubit in a chain or ring into a given state and to
transfer the state, as well as possible, along the chain (or ring)
to the output qubit. The input and output qubits were placed at
diametrically opposite places in the chain (or ring) and the
interaction between neighbouring qubits was time-independent and
constant around the length of the system. Interestingly if a ring
has length four, there is a time after which the state of the
output qubit is precisely the same as that of the input qubit. As
the length of the system is increased, however, the maximum
fidelity achievable was found to go down with increasing
separation. For example, chains of length approximately $80$ have
fidelity not much better that $2/3$ (this is the fidelity of the
best classical transmission of an unknown qubit); thus the
approach would not be useful for transmission over long distances.
A further notable feature of the protocol is that the time at
which the maximum fidelity is reached does not depend in any
simple way on the length of the ring.  This is because the maximum
fidelity arises when an expression of the form $\sum_j
e^{i\pi\alpha_j t}$ reaches its maximum (where $t$ is the time,
and the $\alpha_j$ are rational constants which depend on the
length of the ring). Thus in this framework there is no natural
notion of the speed of transmission along the fibre.

Further progress was made in \cite{christandl:2003a} where the
authors show that a single qubit can be communicated
\emph{perfectly} in a hypercube geometry. The authors also show
how to communicate a qubit perfectly in a linear chain with
variable couplings given by a specific arithmetic sequence.  Using
this idea for communicating arbitrary distances, would however
require re-fabricating the channel for each distance.

A further interesting proposal for communicating quantum
information using the ground state of a certain spin system has
been recently developed in \cite{verstraete:2003a}. There it was
found that \emph{perfect} transmission is possible when the
freedom to apply \emph{local} measurements on all the spins is
allowed.

We can summarise our desiderata in this paper by saying that we
wish to develop schemes for \emph{high fidelity} quantum
communication with three essential criteria; (i) Minimum control
requirements. We require that it is unnecessary to apply many fast
control operations throughout the transmission; (ii) Robustness.
We want the device to be capable of tolerating small errors
without diminishing the quality of the communication too much; and
(iii) Flexibility. We want to be able to change parameters (such
as the locations of the communicating parties) simply without
requiring a new device be fabricated.

Our proposal makes quantum spin chains behave as a \emph{medium}
through which quantum information \emph{propagates}. Using the
fact that  dynamics in restricted subspaces for $1$D quantum spin
chains can be described by  classical fourier wave analysis  we
show that there are natural notions of \emph{group velocity} and
\emph{dispersion} for quantum information. We propose a figure of
merit for the information propagation, the \emph{qubit rate}, and
analyse this quantity for a family of ideal and realistic physical
models: nearest-neighbour $1$D spin chains. In addition, we argue
that our protocols satisfy the requirements (i)-(iii) we outlined
in the previous paragraph.

We show how to increase the quality of quantum communication in a
variety of realistic spin models by relaxing the restriction that
the two parties can only access a single qubit. We allow both
parties the ability to access a limited number of extra sites, a
fraction which tends to zero as the length of the chain increases.
The connection between propagation of quantum information in spin
rings and classical fourier wave analysis provides a way to
\emph{visualise} the \emph{propagation} of quantum information as
a pulse through a medium.

We now turn to a precise description of the protocol we will
study. We consider a ring of $N$ spin-$1/2$ systems evolving
according to a nearest-neighbour hamiltonian $H$. We identify the
$(N+1)$th site with the $1$st site, i.e.\ $A_{N+1}\equiv A_1$. We
imagine that two parties, Alice $A$ and Bob $B$ are located at
sites $1$ and $N/2$, respectively. (We assume, for simplicity,
that the ring is composed of an even number of spins. Note,
however, that our subsequent results do not depend on this fact.)
We suppose that Alice and Bob are able to access a total of
$\Lambda$ sites each, centred on the $1$st and $N/2$th sites,
respectively. Alice and Bob are allowed to perform any operation
allowed by the rules of quantum mechanics on their sites. Finally,
we assume that the completely polarised state $|00\cdots0\rangle$
is an eigenstate of the hamiltonian $H$, and that before the
protocol begins the system is prepared in this state.

Alice wants to communicate a (possibly unknown) qubit state
$|\psi\rangle$ to Bob. Given a specification of $H$ (assumed to be
known to both Alice and Bob) Alice performs some \emph{encoding
operation} $U_A$ on her $\Lambda$ qubits. The system is now in the
state $|\Psi(0)\rangle=U_A|00\cdots0\rangle$. As the ring is
always evolving according to $H$ this state immediately begins to
evolve: $|\Psi(t)\rangle = e^{-iHt}|\Psi(0)\rangle$. Bob now waits
a certain period of time $T$ which depends only on $H$, $\Lambda$
and $N$, and when this duration elapses he performs a
\emph{decoding operation} $U_B$ on his addressable qubits in order
to decode or \emph{refocus} the communicated state into one of the
qubits in the ring. (Additionally, Bob could apply a swap
operation to move the decoded state into a static register qubit.
We prefer to ignore such register qubits because we imagine that
decoding unitary operations could also be performed as
intermediate steps in a kind of \emph{quantum repeater}.) The
protocol is deemed to succeed when the \emph{average fidelity}
\begin{equation}\label{eq:avgfid}
\mathcal{F}(H,N,\Lambda,T,U_A,U_B) \triangleq \frac{1}{4\pi}\int
d\Omega \, \langle\psi|\rho|\psi\rangle,
\end{equation}
where $|\psi\rangle$ is the input state, which we average over the
Bloch sphere, and $\rho$ is the decoded output state, is above
some prespecified threshold value $\tau$.

Ultimately we are interested in how well a spin ring performs as a
conduit for quantum information. In this more general scenario we
allow Alice and Bob the freedom to apply a number of
encoding/decoding operations in succession. The objective is to
maximise the number of qubits successfully communicated (i.e.\
when $\mathcal{F}\ge\tau$) per unit time. Write the maximum number
of qubits that can successfully communicated (where the maximum is
taken over the encoding/decoding operations) in a duration $T$ as
$M_\tau(T)$. We define the long-time average $Q_\tau(H)\triangleq
\lim_{T\rightarrow\infty} \frac{1}{T}M_\tau(T)$ to be the
\emph{qubit rate} for $H$.

Obviously the evaluation of $Q_\tau(H)$ is extremely difficult,
even for the simplest systems. We only obtain lower bounds for
this quantity for a class of rotationally invariant
nearest-neighbour hamiltonians.

There is a simple case where we can evaluate $Q_\tau(H)$ exactly,
namely when $H$ is a hamiltonian corresponding to the translation
operator $H = -i\log(\mathcal{T})$, where $\mathcal{T}$ is defined
by the following action on computational basis vectors,
$\mathcal{T}(|a_1,a_2,\ldots,a_N\rangle) \triangleq |a_2, a_3,
\ldots, a_N, a_1\rangle$. When $H$ is of this form the qubit rate
takes the maximum value $Q_\tau(H)=1$ for all $\tau$. (We have
adopted rescaled time units; we will discuss how to calculate the
constant of rescaling later.) In order to see this first note that
when $t=n$ is an integer the propagator
$U(n)=e^{-iHn}=\mathcal{T}^n$ is a power of the translation
operator. To achieve the qubit rate, Alice needs to encode the
state of a qubit into one site of the ring at every integral $t$.

As we'll argue presently, every hamiltonian $H$ corresponding to
the translation operation is massively nonlocal and contains
interaction terms between many separated subsystems. For this
reason it is unlikely that a system will be fabricated which
naturally evolves according to $H$.

Consider now the translation operator $\mathcal{T}$. It is easy to
see that the action of $\mathcal{T}$ on computational basis states
breaks up into blocks, for example, the states $|00\cdots0\rangle$
and $|11\cdots 1\rangle$ form blocks all by themselves, and the
states $|j\rangle$, $j=1,\ldots,N$, where $|j\rangle$ denotes the
state of all $|0\rangle$'s with a $|1\rangle$ in the $j$th site,
form a block (we'll often refer to these states as
\emph{one-particle states}). The remaining basis states each
segregate into blocks in a similar fashion. Consider a block of
states of size $M$. Choose a state from this block and call it
$|\alpha_1\rangle$. Every state $|\alpha_k\rangle$ within the
block can be written as $|\alpha_k\rangle =
\mathcal{T}^{k-1}|\alpha_0\rangle$. Using these $M$ states one can
construct $M$ eigenstates of $\mathcal{T}$: $|\beta_k\rangle =
1/\sqrt{M}\sum_{j=1}^M \nu^{(j-1)k}|\alpha_j\rangle$, where $\nu =
e^{\frac{2\pi}{M}i}$ is the $M$th root of unity. We note that each
set of eigenstates so constructed is a quantum fourier transform
of the set $\{|\alpha_j\rangle\,|\,j=1,\ldots,M\}$. Performing
this procedure for each block gives rise to the complete set of
eigenstates for $\mathcal{T}$.

We study the qubit rate for the class of rotationally invariant
nearest-neighbour spin-$1/2$ models. Because of the rotational
invariance $[H,\mathcal{T}] = 0$ and we may simultaneously
diagonalise both $H$ and $\mathcal{T}$. We restrict our attention
further to the class of $1$D spin rings which fix the following
special eigenstates of $\mathcal{T}$, which we call the
\emph{twisted $W$-states}, $|W(k)\rangle \triangleq
\frac{1}{\sqrt{N}}\sum_{j=1}^N \mu^{(j-1)k}|j\rangle$, where $\mu$
is the $N$th root of unity $\mu=e^{\frac{2\pi}{N}i}$. All the
protocols we describe in this paper take place in the subspace
formed by the twisted $W$-states.

The spin hamiltonian $H$ fixes the states $|W(k)\rangle$ with
eigenvalues $\omega(k)$ (and also, via a rescaling of energy, $H$
fixes $|00\cdots0\rangle$ with eigenvalue $0$). Suppose we start
the system in the arbitrary state $|\Psi(0)\rangle
=\alpha|00\cdots0\rangle + \beta\sum_{j=1}^N c_j|j\rangle$, where
$|\alpha|^2+|\beta|^2=1$ and $\sum_{j=1}^N |c_j|^2 =1$. The
subsequent time evolution of this state can be written
\begin{equation}
|\Psi(t)\rangle = \alpha|00\cdots0\rangle + \beta\sum_{k=1}^N
\tilde{c}_ke^{-i\omega(k)t}|W(k)\rangle,
\end{equation}
where $\tilde{c}_k=\frac{1}{\sqrt{N}}\sum_{j=1}^N
{\mu^{-(j-1)k}c_j}$. The coefficient $c_j(t)$ of the $|j\rangle$
term can be found as $$c_j(t) = \frac{1}{\sqrt{N}}\sum_{k=1}^N
\tilde{c}_ke^{-i\omega(k)t+2\pi i (j-1)k/N}.$$

Because the dynamics only take place in the zero- and one-particle
subspace we can establish the following useful result. Suppose
$\alpha=0$. If we bipartition the spin system into two subsystems
$A$ and $\overline{A}$ we can always write the system's state
$|\Psi(t)\rangle = \sum_{j=1}^N e_j(t)|j\rangle$ in the following
way,
\begin{equation}\label{eq:schmidtw}
|\Psi(t)\rangle =
\sqrt{1-\mathcal{C}_{\overline{A}}(t)}|\phi\rangle_A|\mathbf{0}\rangle_{\overline{A}}
+
\sqrt{\mathcal{C}_{\overline{A}}(t)}|\mathbf{0}\rangle_A|\phi'\rangle_{\overline{A}},
\end{equation}
where $$\mathcal{C}_{\overline{A}}(t) = \sum_{j\in \overline{A}}
|e_j(t)|^2,$$
$$|\phi\rangle=\frac{1}{\sqrt{1-\mathcal{C}_{\overline{A}}(t)}}\sum_{j\in
A} e_j(t)|j\rangle,$$ and
$$|\phi'\rangle=\frac{1}{\sqrt{\mathcal{C}_{\overline{A}}(t)}}\sum_{j\in
\overline{A}} e_j(t)|j\rangle.$$ Note that Eq.~(\ref{eq:schmidtw})
is a two-term Schmidt decomposition of $|\Psi(t)\rangle$ because
${}_A\langle\phi|\mathbf{0}\rangle_A =
{}_{\overline{A}}\langle\phi'|\mathbf{0}\rangle_{\overline{A}} =
0$.

When Alice wants to send the state $|\psi\rangle = \alpha|0\rangle
+ \beta|1\rangle$ she will encode this state as
$\alpha|00\cdots0\rangle + \beta\sum_{j=1}^N c_j(0)|j\rangle$,
where she is free to choose nonzero $c_j(0)$ as long as the index
$j$ lies within her subset of addressable spins. After time $T$
has elapsed then this state can be written, following the
discussion in the previous paragraph, as
\begin{multline}
|\Psi(T)\rangle =
\beta\sqrt{1-\mathcal{C}_B(T)}|\eta\rangle_{\overline{B}}|\mathbf{0}\rangle_B
+\\
|\mathbf{0}\rangle_{\overline{B}}(\alpha|\mathbf{0}\rangle_B+\beta\sqrt{\mathcal{C}_B(T)}|\eta'\rangle_B),
\end{multline}
where now the bipartition is given by $\overline{B}B$, where $B$
is Bob's addressable spins and $\overline{B}$ now refers to all
the other spins and
$$ \mathcal{C}_B(T) = \sum_{j\in B}
|e_j(t)|^2.$$

Bob now applies a decoding unitary $U_B$ to his part $B$ of the
ring. The unitary is (partially) defined by
$U_B|\mathbf{0}\rangle_B = |\mathbf{0}_B\rangle$ and
$U_B|\eta'\rangle_B = |{N/2}\rangle$. (There is a great deal of
arbitrariness in how Bob decodes his state. We choose to
concentrate the state into one qubit in order to facilitate the
evaluation of the average fidelity of the channel.)

After decoding, the state $\rho$ of the qubit $N/2$ has density
operator
\begin{equation}
\rho = \left( \begin{matrix} |\alpha|^2+
|\beta|^2(1-\mathcal{C}_B(T))  &
\sqrt{\mathcal{C}_B(T)}\alpha\beta^* \\
\sqrt{\mathcal{C}_B(T)}\alpha^*\beta & |\beta|^2\mathcal{C}_B(T)
 \end{matrix} \right)
\end{equation}
with respect to the basis $\{|0\rangle_{N/2}, |1\rangle_{N/2}\}$
of qubit $N/2$. This state may be written
$\rho=\mathcal{E}_T(|\psi\rangle_{N/2}\langle\psi|)=M_0|\psi\rangle_{N/2}\langle\psi|M_0^\dag
+ M_1|\psi\rangle_{N/2}\langle\psi|M_1^\dag$, where
$|\psi\rangle_{N/2} = \alpha|0\rangle_{N/2} +
\beta|1\rangle_{N/2}$ and $M_0$ and $M_1$ are the Kraus operators
of an amplitude damping channel $\mathcal{E}$,
$$M_0 = \left(\begin{matrix} 1 & 0\\ 0 &
\sqrt{\mathcal{C}_B(T)}\end{matrix}\right)$$ and $$M_1 =
\left(\begin{matrix} 0 & \sqrt{1-\mathcal{C}_B(T)}\\ 0 &
0\end{matrix}\right),$$ with respect to the basis
$\{|0\rangle_{N/2}, |1\rangle_{N/2}\}$ (cf.\ \cite{bose:2002a}).

The fidelity $\langle \psi|U_B\rho U_B^\dag |\psi\rangle$ of Bob's
state with the input state is given by
\begin{equation}
|\alpha|^4 +
(1+2\sqrt{\mathcal{C}_B(T)}-\mathcal{C}_B(T))|\alpha|^2|\beta|^2 +
\mathcal{C}_B(T)|\beta|^4.
\end{equation}
The average fidelity Eq.~(\ref{eq:avgfid}) can be evaluated as
$\frac{1}{2} +\frac{1}{3}\sqrt{\mathcal{C}_B(T)}+
\frac{1}{6}\mathcal{C}_B(T)$. This function depends monotonically
on the quantity $\mathcal{C}_B(T)$.

It is clear from the discussion in the preceding paragraphs that
the quantity $\mathcal{C}_B(T)$ plays a central role in
determining the effectiveness of the communication protocol. We
now study this quantity further, and show how to reduce the
problem of its maximisation to a well-known type of problem in
Fourier wave analysis.

Motivated by the importance of $\mathcal{C}_B(T)$ we introduce a
method for visualising quantum states which lie within the
one-particle subspace. Given such a state $|\nu\rangle =
\alpha|00\cdots0\rangle + \beta\sum_{j=1}^N c_j|j\rangle$, we
visualise it by plotting the quantities $\nu_j\triangleq|c_j|^2$
against site number. Clearly the area under the resulting (bar)
graph equals $1$. Obviously this pictorial representation does not
represent any of the phase information contained in the system's
state. However, this phase information plays no part in the
quantity we are interested in, $\mathcal{C}_B(T)$, which is simply
the area under the graph of the state $|\nu(T)\rangle$ between
sites $N/2-\Lambda/2,\ldots,N/2+\Lambda/2$. The graph of the
system's state $|\nu(T)\rangle$ is given by $$\nu_j(T)
=\frac{1}{N}\left|\sum_{k=1}^N \tilde{c}_ke^{-i\omega(k)t+2\pi
i(j-1)k/N}\right|^2.$$

It is worth emphasising that the method we have just introduced
for visualising quantum states of the ring depends crucially on
the properties of our spin systems. The dynamics for the ring
occurs solely within the zero- and one-particle subspaces spanned
by $|\mathbf{0}\rangle$ and $\{|j\rangle\,|\,j=1,\ldots,N\}$. This
subspace scales \emph{linearly} with the ring. It is precisely
this feature which facilitates our construction for the graph of
the state, which is essentially a method for representing,
visually, a linear number (in $N$) of degrees of freedom.

The problem of maximising the area under the graph $\nu_j(T)$
between certain sites is a well-known a classical problem in
fourier wave analysis; our problem has reduced to studying the
linear dynamics of superpositions of complex \emph{scalar} waves.
The quantity $\omega(k)$ is the \emph{dispersion relation} and the
\emph{position variable} is given by $x_j = (j-1)/N$. We wish to
work out an assignment $c_j(0)$ of initial amplitudes in Alice's
subsystem, or a \emph{wavepacket} of quantum information, so that
subsequent evolution preserves as much of the width and integrity
of the wavepacket as possible.

Given a specification of the dispersion relation $\omega(k)$ we
have all the information we require to embark on the design of
wavepackets which preserve their shape. These numbers are, of
course, the eigenvalues of $H$ for the eigenstates $|W(k)\rangle$.

To illustrate our ideas in the following we use the Heisenberg
model on $N$ sites $$H=\chi\frac{N}{2}I-\frac{\chi}{2}\sum_{j=1}^N
\boldsymbol{\sigma}_j\cdot \boldsymbol{\sigma}_{j+1},$$ where we
have introduced an arbitrary constant $\chi$. We will explain
which value of $\chi$ we choose in the following. The dispersion
relation for the Heisenberg model is $\omega(k)=
2\chi\left(1-\cos(\frac{2\pi k}{N})\right)$. (We have rescaled the
zero of energy for the Heisenberg model so that
$|\mathbf{0}\rangle$ has eigenvalue zero.)

Currently our problem is not identical to the well-studied
problems in wave motion (see, for example, \cite{whitham:2000a}).
The difference is that we are dealing with wave motion in a
\emph{discrete system} whereas most of the theory of wave motion
(at least in regards to the design of optimal wavepackets) is for
\emph{continuous systems}. In the subsequent discussion we will
make use of the results from the continuous theory, but it must be
understood that these results only hold \emph{asymptotically} in
the limit where $N\rightarrow\infty$ and the intersite spacing
$\delta\rightarrow0$. In particular, we emphasise that the
appearance of \emph{derivatives} must be understood as a formal
device and really only make sense in this limit. We'll justify the
applicability of these asymptotic results presently by comparing
them with numerical results.

In classical fourier wave theory the concept of \emph{group
velocity} plays a crucial role in the design of wavepackets which
preserve their shape as time evolves \cite{whitham:2000a}. The
group velocity $v(k_0)$ for a wavenumber $k_0$ is defined to be
$v(k_0)=\frac{1}{2\pi}\frac{d\omega(k)}{dk}\big|_{k=k_0}$. (For
the models we consider we are given a specification of $\omega(k)$
which makes sense for fractional $k$, so that it is possible to
compute this derivative.) The significance of this quantity is
that an initial wavepacket $\sum_{k=1}^Nw_ke^{2\pi ix_jk}$
consisting of fundamental waves with large amplitude focussed on
wavenumber $k_0$ will translate with velocity $v(k_0)$ as time
evolves.

Unless the system is dispersionless ($\omega(k) = ck$) the initial
wavepacket will spread as time evolves \cite{whitham:2000a}. The
rate-of-spread of the initial wavepacket is proportional to the
second order term
$\frac{1}{2\pi}\frac{d^2\omega(k)}{dk^2}\big|_{k=k_0}$. If this
derivative is zero (as will be the case in one of the models we
consider) there is no appreciable spread of the wavepacket to
second order in $\omega(k)$. There may, however, be contributions
from third and higher order terms, which cause an initial
wavepacket to spread.

The generic strategy \cite{jackson:1999a} for designing
wavepackets which change shape as little as possible is to use a
gaussian-modulated wave of variance $\Delta$ centred at position
$x_k$ and wavenumber $k_0$ corresponding to the maximum available
group velocity $v(k_0)$. In our case the ``wave'' is a twisted
$W$-state, so this ``wavepacket'' is written $|\Psi(0)\rangle =
\alpha|\mathbf{0}\rangle + \beta|G(x_k,k_0,\Delta)\rangle$, where
$$|G(x_k,k_0,\Delta)\rangle= 1/\sqrt{\mathcal{N}}\sum_{j=1}^N
e^{-\frac{(x_j-x_k)^2}{2\Delta^2}+2\pi ik_0 x_j}|j\rangle$$ and
$\mathcal{N}$ is chosen to normalise the state. The graph for the
initial state is given by $$\Psi_j(0) =
\frac{1}{\mathcal{N}}\left|\sum_{j=1}^N
e^{-\frac{(x_j-x_i)^2}{2\Delta^2} + 2\pi ik_0x_j}\right|^2.$$ This
state has exactly the form of a gaussian-modulated complex scalar
wave centred around wavenumber $k_0$ and with (spatial) width $L$.
(We define \emph{width} $L$ to be when more than some prespecified
area, say 95\%, lies within an interval $L$ centred on the maximum
of the gaussian. We note that $L$ is some multiple of $\Delta$.
Throughout the remainder of this paper we choose $L=4\Delta$.)

It is known that an initial gaussian wavepacket, under
time-evolution, remains, to good approximation, a gaussian of
width $L(t)$ satisfying \cite{jackson:1999a}
\begin{equation}\label{eq:gauspread}
\frac{L(t)}{L(0)} = \left[ 1 +
\left(\frac{\omega''(k_0)t}{L^2(0)}\right)^2\right]^{\frac{1}{2}}.
\end{equation}
Because Alice is assumed able to address only $NL(0)$ sites
(recall we have rescaled the length of the ring to be $1$ by
defining the position variable $x_j = (j-1)/N$) we propose using
the gaussian initial pulse introduced in the previous paragraph
truncated (and renormalised) after $NL(0)$ sites.

\begin{figure}
\begin{center}
\includegraphics{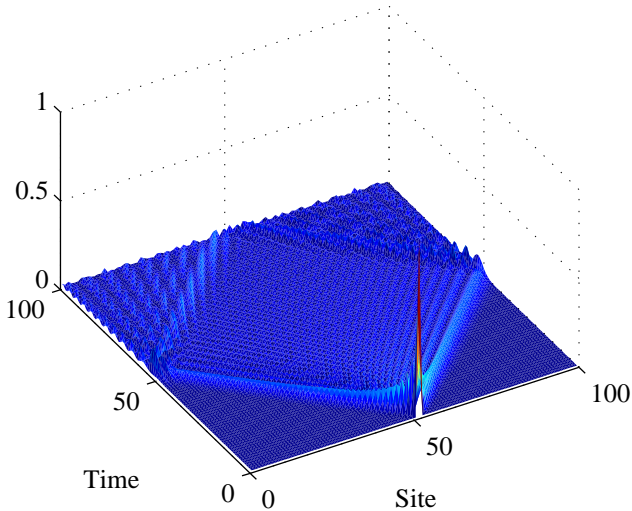}
\caption{Propagation of the state $|N/2\rangle$ in a $100$-site
Heisenberg spin ring.}\label{fig:point}
\end{center}
\end{figure}

\begin{figure}
\begin{center}
\includegraphics{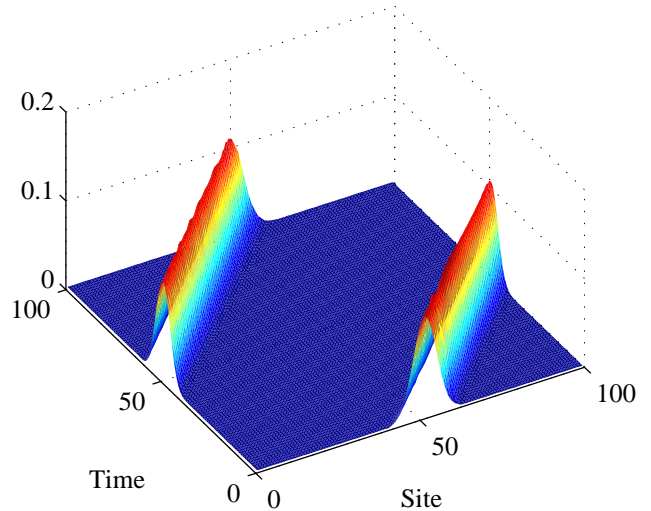}
\caption{Propagation of a truncated gaussian-modulated $W$-state
of width $10$ and wavenumber $N/4$ in a $100$-site Heisenberg spin
ring.}\label{fig:shaped}
\end{center}
\end{figure}

\begin{figure}
\begin{center}
\includegraphics{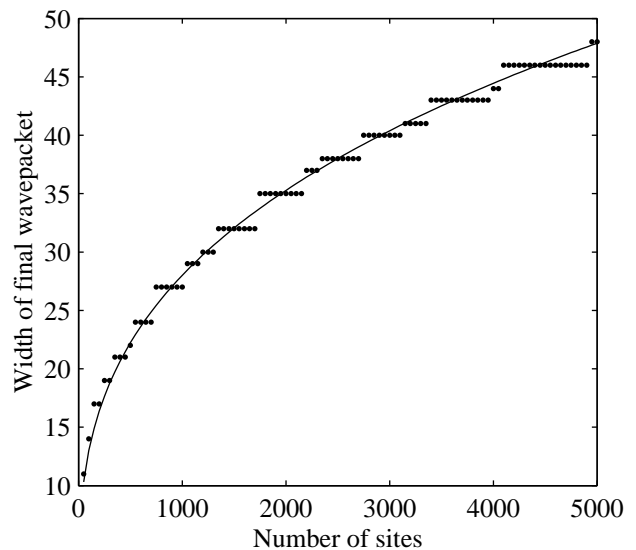}
\caption{Dots: size of the final wavepacket in the Heisenberg spin
ring for $N=50$ to $N=5000$. (Initial wavepacket is a gaussian
modulated wave of variance ${N}^{\frac{1}{3}}$ truncated after
$2{N}^{\frac{1}{3}}$ sites.) Final width $F$ is defined to be when
more than $95\%$ of the wavepacket lies between $F$ sites. Solid
line: graph of $2.8{N}^{\frac{1}{3}}$ for comparison. The apparent
jumps of integer amounts greater than $1$ in the numerical results
is an artifact of the way the width of the final wavepacket is
calculated. The final wavepacket has a sequence trailing
oscillations. In order to get an area of more than some arbitrary
amount (like $0.95$) it is sometimes necessary to include several
more sites in a go.}\label{fig:scaling}
\end{center}
\end{figure}

We want to design a wavepacket that spreads by an amount that
scales favourably with $N$. We demand that a wavepacket of width
$L$ spreads by only a constant factor to a final width of
$L'\approx \kappa L$ by the time it reaches Bob. To simplify the
discussion we note that for typical wavenumbers close to $k_0$ we
have $\omega''(k_0)=\lambda/N^2$ (this is true of all the models
we consider). We analyse the scaling, with $t\approx N$ (the time
it takes for a wavepacket to traverse the ring), of the
\emph{spread} $S(N)$ of the wavepacket: $S(N)=L(N)/L(0) = \left[ 1
+ \left(\frac{\lambda}{NL^2(0)}\right)^2\right]^{\frac{1}{2}}$. If
we want the final width of the wavepacket to be independent of the
length of the ring we need to choose $L(0)=1/\sqrt{N}$. This means
that the initial packet needs to be supported on a subsystem
containing on the order of $\sqrt{N}$ sites. In the ``mesoscopic''
limit of large but finite $N$ the fraction of the ring that Alice
and Bob need to access can be made arbitrarily small (it scales as
$N^{-\frac{1}{2}}$).

We now analyse these results for the Heisenberg model. We want a
signal pulse to travel as fast as possible. Looking at the group
velocity $\omega'(k) = \frac{2\chi}{N} \sin(\frac{2\pi}{N}k)$ we
find that this occurs for $k=N/4,3N/4$, $\nu(k) =
\pm\frac{2\chi}{N}$. At this point we choose $\chi=1/4$. We do
this so that the final wavepacket will arrive at Bob's portion of
the ring at time $t=N$ (recall that Alice and Bob are separated by
a distance $N/2$). So we choose the initial wavepacket to be a
truncated gaussian-modulated superposition of width $\sqrt{N}$ of
twisted $W$-states centred around $k=N/4$.

The case of wavepackets centred on $k_0=N/4$ is rather special ---
note that the dispersion $\omega''(k_0)$ for this model is
identically $0$. Na{\"\i}vely applying formula
Eq.~(\ref{eq:gauspread}) for the spread of the wavepacket suggests
that an initial wavepacket of \emph{any} width will not spread at
all. This is  an artifact of our approximation; we assumed that
the dispersion relation could be expanded and truncated at second
order with little error. In the special situation where
$\omega''(k_0)$ vanishes we need to go to third order. For this
case a gaussian pulse does not remain a gaussian. (Instead, it
becomes an Airy function \cite{agrawal:2002a}.) The formula
Eq.~(\ref{eq:gauspread}) is invalid in this case and we need to
use the \emph{third-order broadening factor} \cite{agrawal:2002a}
$$ \frac{L(t)}{L(0)} =
\left[1+\frac12\left(\frac{\omega'''(k_0)t}{\sqrt{2}L^3(0)}\right)^2\right]^{\frac12}.$$
We solve this equation for $t=N$ to find the width $L(0)$ of
initial wavepacket which has constant spread with $N$. We find
that $L(0)$ need only scale as $N^{-\frac23}$. Such a wavepacket
consists of $NL(0)=N^{\frac{1}{3}}$ sites.

We illustrate these results for the Heisenberg ring of $N=100$
sites in Fig.~(\ref{fig:point}) and Fig.~(\ref{fig:shaped}). By
way of contrast, in Fig.~(\ref{fig:point}) we first show the
dynamics of the state with a single $|1\rangle$ at site $N/2$
(this type of localised initial state is what is used in the
protocols of \cite{bose:2002a}, \cite{subrahmanyam:2003a} and
\cite{christandl:2003a}). Because the state $|N/2\rangle$ is an
equal superposition of \emph{all} the twisted $W$-states all the
waves are involved in the dynamics and hence the initial packet
disperses rapidly. In Fig.~(\ref{fig:shaped}) we see that the
shaped pulse retains its shape for much longer durations. This is
because fewer waves are involved in the superposition. We note
that, generically, for wavepackets centred around wavenumbers not
equal to $N/4$ a square-root scaling for the final wavepacket is
observed.

Before we conclude our discussion of the numerical results we
refer to Fig.~(\ref{fig:scaling}). In this graph we have plotted,
for the Heisenberg ring, the width of the final wavepacket (we
define the width $F$ to be when more than $95\%$ of the area lies
within $F/2$ sites either side of site $N/2$) for rings of sizes
$50$ through $5000$ sites. We have also plotted
$2.8{N}^{\frac{1}{3}}$ for comparison. Because the Heisenberg
model is diffusionless to second order in the dispersion relation
$\omega''(k_0) = 0$, we expect that the numerically recorded
spread ought to be smaller than the spread predicted for models
with second-order dispersion. We can see, from
Fig.~(\ref{fig:scaling}) that this is indeed the case. For
$N=5000$ the proportion of the ring that Alice and Bob must be
able to access is less than $1\%$ each.

To conclude we construct all the rotationally invariant
nearest-neighbour spin hamiltonians which fix the
$|\mathbf{0}\rangle$ and twisted $W$-states. An easy way to
construct all such hamiltonians is to first assume that the
hamiltonian preserves $S^z \triangleq \sum_{j=1}^N \sigma_j^z$.
Any such hamiltonian may be written as
\begin{multline}\label{eq:genrotham}
H = c_0 I + \sum_{j=1}^{N} \Big[ c_1\sigma^+_j \sigma^-_j +
c_2\sigma^-_j \sigma^+_j + d_1(\sigma^+_j \sigma^-_{j+1} +
\sigma^-_j \sigma^+_{j+1}) \\ +d_2(\sigma^+_j \sigma^-_{j+1} +
\sigma^-_j \sigma^+_{j+1})^2 + e_1i(\sigma^+_j \sigma^-_{j+1} -
\sigma^-_j \sigma^+_{j+1})\\ +f_1 i(\sigma^+_j \sigma^-_{j+1} -
\sigma^-_j \sigma^+_{j+1})(\sigma^+_j \sigma^-_{j+1} + \sigma^-_j
\sigma^+_{j+1}) \Big],
\end{multline}
where $\sigma^{\pm}_j \triangleq \frac{1}{2}(\sigma^x_j\pm
i\sigma^y_j)$. It may be verified, with a little algebra, that
this is the most general rotationally invariant nearest-neighbour
hamiltonian which nontrivially fixes the twisted $W$-states.
(There is one additional term which fixes the twisted $W$-states,
$\sigma^-_j \sigma^-_{j+1}$, however, it trivially annihilates
these states and thus contributes nothing to the dispersion
relation. Note also that the $f_1$ term annihilates the twisted
$W$-states.)

The hamiltonian Eq.~(\ref{eq:genrotham}) gives rise to the
following dispersion relation
\begin{equation}\label{eq:nndisprel}
\omega(k) = A + B \cos\left(\frac{2\pi}{N}k\right) +
B'\sin\left(\frac{2\pi}{N}k\right),
\end{equation}
where $A = c_0 + c_1(N-1) + c_2 + 2d_2$, $B = 2d_1$ and $B' =
-2e_1$. We note that both the Heisenberg model in a magnetic field
and the $XY$ model in a magnetic field can both be expressed as in
Eq.~(\ref{eq:genrotham}) for specific choices of $c_j$, $d_j$,
$e_1$ and $f_1$. The dispersion relation Eq.~(\ref{eq:nndisprel})
is no more general than that for the Heisenberg model so all of
our discussion concerning the design of optimal signals for the
Heisenberg model carries through straightforwardly for the general
class of models Eq.~(\ref{eq:genrotham}).

We see from the dispersion relation Eq.~(\ref{eq:nndisprel}) that
the most general rotationally invariant nearest-neighbour
hamiltonian fixing the twisted $W$-states is dispersive. This
shows us that, in order for a hamiltonian $H$ to exponentiate to
the translation operator $\mathcal{T}=e^{iH}$, it is necessary for
$H$ to contain interaction terms between separated parties. This
verifies our earlier claim that such a hamiltonian must be
nonlocal.

Finally, we note that our results imply that the qubit rate for a
rotationally invariant nearest-neighbour Hamiltonian on $N$ sites
is bounded below: $Q_{0.95}(H)\ge 1/N$.

We have introduced a method for improving the quantum
communication characteristics of $1$D quantum spin rings.
According to the connection between the dynamics of quantum
information in these rings with fourier wave analysis we have been
able to import many of the results concerning the design of
signals which disperse minimally. Clearly our results illustrate
that our communication protocol has minimal control requirements
and is flexible. We have not tackled the problem of determining
the resistance of our protocol to error.

Many future problems suggest themselves at this stage. Perhaps the
most interesting is the extension of the results to linear chains.
In this case our results cannot be applied because the twisted
$W$-states are not eigenstates for nearest-neighbour hamiltonians
on a chain. However, a generalisation of our method of pulse
shaping ought to be possible. Another important future problem
concerns the resistance of the protocol to errors. We believe that
the protocol is robust, but a full analysis needs to be done.
Finally, we note that all our protocols take place in the
single-particle subspace. The dimension of this subspace increases
\emph{linearly} with increasing numbers of qubits; however the
hilbert space of the system increases \emph{exponentially} with
$N$. Perhaps it is possible to increase the qubit rate
substantially by taking advantage of larger subspaces which
include two and higher particles. Further investigations along
these lines are being conducted.

\begin{acknowledgments}
We are grateful to the EU for support for this research under the
IST project RESQ.
\end{acknowledgments}

\end{document}